\documentclass{aastex}
\usepackage{spr-astr-addons}
\usepackage{url}
\usepackage{multicol}

\begin{document}

\title{15-Digit Accuracy Calculations of Chandrasekhar's $H$-function  for Isotropic Scattering\\ by Means of the Double Exponential  Formula}
\slugcomment{Accepted for publication in  Astrophys. Space Sci.}
\shorttitle{Calculations of the $H$-function for Isotropic Scattering }
\shortauthors{Kawabata}

\author{Kiyoshi Kawabata} 
\affil{Department of Physics, Tokyo University of Science,\\
  Shinjuku-ku, Tokyo 162-8601, Japan\\ e-mail: kawabata@rs.kagu.tus.ac.jp}



\begin{abstract}
This work shows that it is possible to calculate numerical values of the Chandrasekhar $H$-function for isotropic scattering at least with 15-digit accuracy 
by making use of   the double exponential  formula~(DE-formula) of Takahashi and Mori (Publ. RIMS, Kyoto Univ. {\bf 9}, 721, 1974) instead of the Gauss-Legendre quadrature employed 
 in the numerical scheme of Kawabata and Limaye (Astrophys. Space Sci. {\bf 332}, 365, 2011) and simultaneously taking a precautionary measure to minimize the effects due to  loss 
 of significant digits particularly in the cases of near-conservative scattering and/or errors  involved in  returned values of library  functions supplied  by  compilers in use. 
The results of our calculations are presented for 18 selected values of single scattering albedo $\varpi_0$ and 22 values of an angular variable $\mu$, the cosine of zenith angle  $\theta$ specifying the direction of  radiation incident on  or emergent from semi-infinite media.
\end{abstract}



\section{Introduction}

We can express  the emergent intensities of radiation reflected by semi-infinite, vertically homogeneous media  in terms of  Chandrasekhar's $H$-functions $H(\varpi_0,\mu)$, the solution of the following integral equation: 
\begin{equation}
\displaystyle{H(\varpi_0,\mu)=1+\mu H(\varpi_0,\mu)\int_0^1\frac{\Psi(\varpi_0,\eta)}{\mu+\eta}H(\varpi_0,\eta)d\eta},  \label{eq-for-H}
\end{equation}
where $\varpi_0$ is the single scattering albedo, $\mu$ is the cosine of  the zenith angle $\theta$ specifying the direction of incident or emergent direction of radiation, and $\Psi(\varpi_0, \eta)$  is the characteristic function representing the type of scattering the radiation undergoes~\citep[][]{chan1960}.  Lately, \citet{kawa2015} developed an efficient iterative scheme to  obtain numerical values of the $H(\varpi_0,\mu)$ with 11-digit accuracy
 by making use of the approximate interpolation formula obtained  by \cite{kl2011}~\citep[see also][for erratum]{kl2013} for the  $H(\varpi_0,\mu)$ for isotropic scattering,  which corresponds to $\Psi(\varpi_0, \eta)=\frac{1}{2}\varpi_0$. \par
 Accurate numerical evaluation of the $H$-function is important not only for radiative transfer-related problems but also for  other disciplinary areas  such as the condensed matter physics and the theory of neutron transport as has been pointed out by, e.g.,  \cite{jab2012}.  
An outstanding work from this standpoint  was made by \cite{viik1986}, who succeeded in calculating  numerical values of the $H$, $X$, and $Y$-functions of Chandrasekhar\footnote{They are often referred to as \textit{Ambarzumian-Chandrasekhar's functions} in Russian literature~\citep[see, e.g., ][]{sobo1975, amba1942, amba1944}.
However, we shall  adhere to the name  \textit{Chandrasekhar's functions}  just for brevity.} with 14-digit accuracy
 approximating Sobolev's resolvent function employing exponent series.  
Equally interesting work  was done  by \cite{mona2007}, who evaluated  the $H$-function for isotropic scattering via $X$-function of neutron transport for single speed and isotropic case together with the application of the  double exponential formula (DE-formula) proposed by \cite{tm1974}.\par 
However, in the case of the $H$-function for isotropic scattering, various forms of integral representations such as the one employed by \cite{kl2011} are known.   It would therefore be interesting to investigate if it is possible to calculate the values of the isotropic scattering $H$-function using  an integral representation with accuracy comparable to that of \cite{viik1986}.

\section{Formalism}

Since our main interest is  in applications to radiative transfer problems, we may assume that the parameter $\varpi_0$ lies in the range $[0, 1]$.
As in \cite{kl2011}, let us express the $H$-function for isotropic scattering in a closed form:
\begin{equation}
\displaystyle{H(\varpi_0, \mu)=\exp\left[-(\mu/\pi)\int_0^{\pi/2}\!\!f_1(\varpi_0, x)f_2(\mu, x)dx\right], }\label{eq-for-Hiso}
\end{equation}
where we have defined the functions $f_1$ and $f_2$ as
\begin{subequations}
\begin{eqnarray}
f_1(\varpi_0, x)&=&      \ln(1-\varpi_0 x /\tan x),  \label{f1}\\
f_2(\mu, x)&=&  (1+\tan^2 x)/\left[1+(\mu\tan x)^2\right].\label{f2}
\end{eqnarray}
\end{subequations}
Evidently,  $H(0, \mu)=1$ holds irrespective of the values of $\mu$ due to the fact that $f_1(0, x)=0$.\par
It should be noted that \cite{kl2011} applied the Gauss-Legendre quadrature to perform  the integration  over the interval $[0, \pi/2]$  in Eq.\eqref{eq-for-Hiso} for non-conservative cases, viz., $\varpi_0<1$. In the conservative case, on the other hand,  they partitioned the  domain   into $[0, \varepsilon]$ and  $[\varepsilon, \pi/2]$,  and 
an analytical integration was carried out for the first  interval  assuming $\varepsilon\ll 1$, thereby avoiding the numerical difficulty that would arise from   a logarithmic divergence of the integrand at $x=0$, while the Gauss-Legendre  quadrature was applied to the second interval.   \par
The most straightforward way to fulfill  our objective would therefore be to replace the Gauss-Legendre  quadrature 
with something superior.  This  premise and the work of \cite{mona2007}   naturally led us  to trying  the  \textit{double exponential formula} (DE-formula) of \cite{tm1974} for numerical integrations.  The optimality of the DE-formula has been mathematically proven by \cite{sug1997}, indicating  that  it is capable of giving the most accurate result by the minimum number of function evaluations \citep[][]{ms2001}. The method is known to be efficient even for integrals with end-point singularities. Furthermore, it can easily be incorporated with  integrations equipped with  automatic step-size adjustment. What is more,  halving the step-size approximately doubles the number of significant figures~\citep[][]{tm1974}.
Obviously, therefore, the DE-formula is  a highly   promising  alternative to the Gauss-Legendre quadrature employed by \cite{kl2011}. \par
  In writing our  automatic integrator in  FORTRAN code applying  the DE-formula to calculations of the $H$-function, a recourse has been  made to the  subroutine \textit{DEAUTO}, a RATFOR code,  published  by \cite{wat1990}, which ingeniously circumvents the overflow problem often encountered when computing the weights of the DE-formula ~\citep[see also][]{mor1990}. 
  We shall perform all our  calculations in  double-precision arithmetic with the Compaq Visual Fortran compiler Ver. 6.6 for 32-bit computers.\par
In actual  numerical calculations, we closely follow the procedure of \cite{kl2011}: for the conservative scattering case ($\varpi_0=1$), 
the integral involved in Eq.\eqref{eq-for-Hiso} is evaluated in two parts as has been mentioned above:
\begin{align}
H^\text{DE}(\varpi_0, \mu)&=\displaystyle{\exp\left[-(\mu/\pi)\int_0^{\pi/2}\!\!f_1(\varpi_0, x)f_2(\mu, x)dx\right]} \notag \\
&=\exp\left[-(\mu/\pi)(I_1+I_2)\right],\label{HDE}
\end{align}
where we have  
\begin{eqnarray}
I_1&=&\displaystyle{ \int_0^\varepsilon f_1(\varpi_0, x)f_2(\mu, x)
dx } \nonumber  \nonumber \\
&\simeq&\left\{(2\ln\,\varepsilon-2-\ln\,3)+{1\over 45} \left[30(1-\mu^2)\ln\,\varepsilon  \right.  \right. \nonumber \\
& &\left.+5\mu^2(2+3\ln\,3)-3(3+5\ln\,3)\right]\varepsilon^2  \nonumber \\
& &-\left[{617\over 15750}+{2\ln\,3\over 15}-{1\over 75}\mu^2(9+25\ln\,3)\right.  \nonumber \\
& &+{1\over 25}\mu^4(2+5\ln\,3)-{2\over 15}(2-5\mu^2+3\mu^4)\times \nonumber \\
& &\times\ln\,\varepsilon\biggr]\varepsilon^4\biggr\}\varepsilon   \label{eq-I1-integ}
 \end{eqnarray} 
 according to Eq.(A.1) of \cite{kl2011},
and
\begin{equation}
I_2=\displaystyle{\int_\varepsilon^{\pi/2}\!\! f_1(\varpi_0, x)f_2(\mu, x)dx} \label{eq-I2-integ}
\end{equation}
which we shall calculate using an  automatic integrator code based on the DE-formula, and hence the suffix \text{DE} attached to $H(\varpi_0, \mu)$ appearing on the left-hand side of  Eq.\eqref{HDE}.  
The use of Eq.\eqref{eq-I1-integ} enables us to avoid the numerical difficulty arising  from the logarithmic divergence of $f_1(1, x)$ as we approach $x=0$. \par
In the non-conservative cases ($\varpi_0<0$), on the other hand, no such singularity problem occurs  in  the integrand of Eq.\eqref{HDE}, so that the DE-formula is applied to the entire domain  $[0, \pi/2]$ with $\varepsilon$ being set to $0$ in Eq.\eqref{eq-I2-integ}. \par
Furthermore, to minimize the effect of  an error originating  from $f_1(\varpi_0, \mu)$ for small values of $x$ and that due to  loss of significant figures caused when the value of  $\varpi_0$ is close to but not equal  to unity, we shall use the  Maclaurin series expansion truncated at the $x^{12}$ order term: 
\begin{eqnarray}
1-\varpi_0 x/\tan x&\simeq&
\displaystyle{\delta+\frac{\varpi_0x^2}{3}\left[1+\frac{x^2}{15}\left[1+\frac{x^2}{21}\biggl[2+\right.\right. } \nonumber \\
& & \displaystyle{\left.\left.\left.+x^2\left[\frac{1}{5}+\frac{2x^2}{99}\left[1+\frac{691x^2}{6825}\right]\right] \right]\right]\right],}  \nonumber \\
& &  \qquad \qquad (x\le x_\text{t}<1),
\label{inside-of-ln}
\end{eqnarray}
where $x_\text{t}$ is a certain demarcation point and $\varpi_0=1-\delta$ with $\delta$ being  externally specified. Even for $x>x_\text{t}$, 
the use of the following form  has  been found effective for  preventing  loss of significant figures:
\begin{equation}
1-\varpi_0 x/\tan x=1-u+u\delta,   \label{rewrite}
\end{equation}
where we have put  $u=x/\tan x$.\par
 An adequate  set of numerical values for the parameters $\varepsilon$ and $x_\text{t}$ will be determined  applying a modified Powell's method for function minimization \citep[][]{pres1992} as we shall discuss later.  One great advantage of this method is that it does not require gradient of the target function to be minimized. \par
To check whether or not our values of $H$-function have an accuracy of  at least  15 decimal  figures,  we employ an   alternative form of the definition equation Eq.\eqref{eq-for-H} \citep[see Eq.(13) on p.107 of ][]{chan1960}, and perform a single iteration:
\begin{align}
H^\text{It}(\varpi_0, \mu)&=\displaystyle{\left[\sqrt{\delta}+\frac{\varpi_0}{2}\int_0^1\!\!\frac{x H^\text{DE}(\varpi_0, x)}{\mu+x}dx\right]^{-1}, }\nonumber \\
&  \nonumber \\
& \qquad (\varpi_0=1-\delta),  \label{check-for-H}
\end{align}
to polish the values of  $H^\text{DE}(\varpi_0, \mu)$ to get a  refined set of  values $H^\text{It}(\varpi_0, \mu)$.
Needless to say, both $H^\text{It}(\varpi_0, \mu)$  and  $H^\text{DE}(\varpi_0, \mu)$ must   be in agreement with each other to 
  15 digits or more provided that  the latter actually approximates  the true solution for the $H$-function with a 15-digit accuracy, and that the integration over $x$ in Eq.\eqref{check-for-H} has been  done with a comparable accuracy by means of the DE-formula.
An estimation of the numerical error in each  value of  $H^\text{DE}(\varpi_0, \mu)$ may then be made by calculating the difference of the two:
\begin{equation}
\Delta=\left[H^\text{DE}(\varpi_0, \mu)-H^\text{It}(\varpi_0, \mu)\right]\times 10^{15}.  \label{err-of-H}
\end{equation}
Although not explicitly indicated here for brevity, the quantity $\Delta$ depends not only on $\varpi_0$ and $\mu$ but also on 
the  values of $\varepsilon$ and $x_\text{t}$  assumed for the $H$-function calculations.
The multiplicative factor $10^{15}$  is to stress  how many units of difference exist  in the figure at the 15-th decimal place
 of  a given value of $H^\text{DE}(\varpi_0, \mu)$
compared  with  that  of $H^\text{It}(\varpi_0, \mu)$, presumably a closer  approximation   for the true value.  \par
 For the purpose of an additional check, we shall also calculate the zeroth through fourth order moments $\alpha_m(\varpi_0)\ (m=0,\cdots,4)$ of the $H$-function:
\begin{align}
\alpha_m(\varpi_0)&=\displaystyle{\int_0^1\!\!H^\text{DE}(\varpi_0, x) x^m dx, } \nonumber \\
  &\qquad  (m=0, 1, 2, 3, 4),  \label{mth-moment}
\end{align}
which yields the following analytical relations:
\begin{align}
\alpha_0(\varpi_0) & 
             =\begin{cases}
                (2/\varpi_0)[1-\sqrt{\delta}],  \qquad (\varpi_0=1-\delta), & \\
               (128+\varpi_0(32+\varpi_0(16+\varpi_0(10+&  \\
              \qquad +7\varpi_0) )))/128, \!\!\quad  (\varpi_0\le 10^{-3}), &   
               \end{cases}  \label{alpha-0}
\end{align}    
together with 
\begin{subequations}      
\begin{align}     
\alpha_0(1)&=2,   \label{alpha-10}  \\
\alpha_1(1)&=2/\sqrt{3}=1.154700538379251\cdots, \label{alpha-11}\\
\alpha_2(1)&=2q_\infty/\sqrt{3,}= 0.8203524821491256\cdots,  \label{alpha-12}\\
\alpha_3(1)&=\left(\frac{1}{3}q_\infty^2+\frac{1}{5}\right)\sqrt{3}\nonumber \\ &=0.6378182680315181\cdots,   \label{alpha-13}
\end{align}
\label{alphas}
\end{subequations}
\hspace{-0.25cm}where the second expression of Eq.\eqref{alpha-0} is based on the series expansion of $\sqrt{1-\varpi_0}$ with respect to $\varpi_0$ to  the term of order $\varpi_0^5$, and the quantity $q_\infty$ is the Hopf constant \citep[][]{yano1997} or the Hopf extrapolation length \citep[][]{hul1980},
 whose value $0.71044608959876307\cdots$
is given by \cite{viik1986} up to the 59-th decimal place (see also Appendix of the present work).
The  numerical integrations required for Eq.\eqref{mth-moment} are  to be carried out  also using  the DE-formula to achieve as high an accuracy as possible.\par
\section{Numerical Results}
A foremost  reference of comparison to judge the quality of  our numerical results would be Table XI on p. 125 of \cite{chan1960}.  Unfortunately, however, some of the numerical values 
 presented in this table   
 contain non-negligible errors  in the fourth or fifth decimal figures as has already been noted  by \cite{hiro1994}.  Since his tabulation   fully covers   Table XI but with higher  numerical accuracy, 
  we instead made  a direct comparison of our results with those of \cite{hiro1994} by adopting  all of the 47 $\varpi_0$-values employed by him for  tabulation together with nine additional values, viz., 
  $10^{-3}$,  $1-10^{-5}$, $1-10^{-7}$,  
 $1-10^{-9}$, $1-10^{-10}$, $1-10^{-11}$, $1-10^{-12}$, $1-10^{-13}$, and $1-10^{-14}$ to  inspect  the behavior of our  values of the $H$-function  in the presence of  strong extinction or in the near-conservative cases.  
For each of these, the values of the $H$-function  were  calculated at  22 values of $\mu$ comprising 21 values $0~(0.05)~1$ employed  by \cite{hiro1994} also for his tabulation and one extra  value $0.01$. \par  
 For  a given  set of values for the parameters  $(\varepsilon, x_\text{t})$,  $56\times 22$ absolute values of $\Delta$ were obtained following   Eq.\eqref{err-of-H}, which were  then  
    summed up to get the total value $\Delta^\text{Tot}(\varepsilon, x_\text{t})$:
 \begin{equation}
 \Delta^\text{Tot}(\varepsilon, x_\text{t})=\sum_{\varpi_0, \mu} |\Delta|   \label{deltatot}
 \end{equation}
 It is expected that the  set of values for $(\varepsilon, x_\text{t})$ which  minimizes the value of $\Delta^\text{Tot}(\varepsilon, x_\text{t})$ is optimal.   We made  a  search for such pair 
incorporating the subroutine \textit{powell} in \cite{pres1992}  with our FORTRAN code for $H^\text{DE}(\varpi_0, \mu)$ calculations  and treating  the $\Delta^\text{Tot}(\varepsilon, x_\text{t})$ as the target function,  to get to  the following result: 
\begin{equation}
\begin{split}
\varepsilon\phantom{a}&=6.009587099094104\times 10^{-3}, \\ x_\text{t}&=1.238798627279953\times 10^{-1},
\end{split}
\end{equation}
which we adopted to generate the final version of numerical tables of  $H^\text{DE}(\varpi_0, \mu)$.\par
To start with, our numerical results,  rounded to the fifth decimal place, were compared with those of \cite{hiro1994} shown at  $987(=47\times 21)$ entries in his Table 1: the figures in the two data sets were  in perfect agreement at 966 entries, and one unit differences in the fifth decimal possibly arising from round-off errors were found at  the remaining 21 entries. 
 In contrast, Table XI of \cite{chan1960} exhibits  numerous differences in the fifth or even in the fourth decimal compared to ours: for $\varpi_0=0.7$ and $\mu=0.05$, for instance,  
 we as well as \cite{hiro1994} have $1.06765$,  whereas  \cite{chan1960} gives $1.06780$.
 In view of the fact that our result and Hiroi's are identical in spite of the difference in the adopted computational methods, this  difference of $0.00015$ 
 can   duly  be attributed to a fourth decimal error involved in the value shown  in Chandrasekhar's Table XI.  \par
Tables~\ref{tab-1} through \ref{tab-6} show the values of $H^\text{DE}(\varpi_0, \mu)$  to the  15-th decimal place (16 decimal digits)  as functions of $\varpi_0$ and $\mu$: 18  values of $\varpi_0$, viz., $10^{-3}$, $0.1$, $0.3$, $0.5$, $0.7$, $0.8$, $0.9$, $0.99$, $0.999$, $1-10^{-5}$, $1-10^{-7}$,  
 $1-10^{-9}$, $1-10^{-10}$, $1-10^{-11}$, $1-10^{-12}$, $1-10^{-13}$, $1-10^{-14}$, and $1$,  selected from the 56, and  the 22  aforementioned values of $\mu$, viz., $0$, $0.01$, and $0.05~(0.05)~1$ adopted for our calculations.  For brevity, however, we  have dropped the suffix DE in the 
tabulation. \par
The  value of $\Delta$ associated with 
each value of  $H^\text{DE}(\varpi_0, \mu)$  indicates a possible error present in the 15-th decimal figure.
We see that  the largest magnitude of error is 2,  so  that all the results for $H^\text{DE}(\varpi_0, \mu)$ must be   exact  at least to the 14-th decimal place or to 15 decimal  digits.\footnote{The magnitudes of errors are compiler-dependent.  If we run  the same FORTRAN code using, e.g.,  the GFORTRAN Ver. 4.8.1 for 32-bit computers, we get the maximum error of 4 in the 15-th decimal.  }
Also shown in the bottom section of each table are the  values calculated for  the zeroth through fourth  order moments $\alpha_m(\varpi_0)\ (m=0, \cdots, 4)$ of the $H^\text{DE}(\varpi_0, \mu)$  as  functions of $\varpi_0$. \par
As a second  check, we compared our values for  $H^\text{DE}(\varpi_0, \mu)$ and  $\alpha_m(\varpi_0)\ (m=0,\cdots,4)$ with those tabulated in  \cite{bos1983} to the 10-th decimal place, to confirm that both data sets fully agree with each other if ours are rounded to the 10-th decimal place.
Next, our results  for $\alpha_0(\varpi_0)$ and  $\alpha_m(1)\ (m=1, 2, 3)$ were compared with  those given by Eqs.\eqref{alpha-0} and \eqref{alphas},
to find that they too are in  perfect agreement with each other to the 15-th decimal place.\par
 We should also note that \cite{viik1986} gives in his TABLE II  the   values of the $H$-function for isotropic scattering at 11 values of $\mu$, viz., $0~(0.1)~1$, for 
  two values of $\varpi_0$, viz., $0.5$ and $1$,  to the 13-th decimal place.
 The  values  for  $H^\text{DE}(\varpi_0, \mu)$ agree with his  to the  last figures  except that ours are larger by one unit  in the last  decimal place at $\mu=0.2,~ 0.4, ~0.6$, and $0.7$ in the case of  $\varpi_0=0.5$ due probably to round-off errors.  In addition, all our results for $\alpha_m(\varpi_0) \  (m=0, 1, 2)$ for $\varpi_0=0.5$ and $1$ are also found to agree with those of \cite{viik1986} to the 13-th decimal place; an exception is our value of $\alpha_1(1) $, which is larger by one unit  than Viik's in the 13-th decimal place again due possibly  to a round-off error.\par
It may be worth noting that even with the 350-point Gauss-Legendre quadrature, \cite{kl2011} had some difficulty in getting the value of $\alpha_0(\varpi_0)$ correct to the 10-th decimal place for some values of $\varpi_0$ quite in contrast to the present work where we have experienced no such  problem under  the DE-formula.
\section{Conclusion}
We have developed a numerical scheme to calculate  values of Chandrasekhar's $H$-function for isotropic scattering with 15-digit accuracy.   The important elements of this scheme can be summarized  as following:
\begin{enumerate}
\item[1.] The closed form integral representation  for   the $H$-function is  employed as in \cite{kl2011}.
\item[2.] The Gauss-Legendre  quadrature adopted by Kawabata and Limaye~(2011) is replaced with 
the double exponential formula (DE-formula) of Takahashi and Mori~(1974) that enables us to perform the required integrations with  considerably higher   accuracy.  In creating our FORTRAN code,  we gratefully benefited from  
 the automatic integrator routine \textit{DEAUTO} written by \cite{wat1990} in RATFOR, which ingeniously circumvents the over-flow problem frequently encountered in calculating the quadrature weights. We  retain all his  input parameter values also in  our calculations. 
\item[3.] In the case of conservative scattering ($\varpi_0=1$), an analytical integration based on the Maclaurin series expansion of the integrand is used  as in \cite{kl2011} for the integration over a  small interval  $[0, \varepsilon]\ (\varepsilon\ll 1)$ to reduce   numerical inaccuracy  that is likely to be introduced  by the logarithmic divergence of the integrand, whereas  the DE-formula is applied to the integration over the remaining  interval $[\varepsilon, \pi/2]$.  In the cases of non-conservative scattering $(\varpi_0 < 1)$,  on the other hand, only  the DE-formula is applied to the entire  single interval $[0, \pi/2]$.
\item[4.] In the interval $[0, x_\text{t}]\ (x_\text{t}< 1)$, where $x_\text{t}$ is a prescribed  demarcation point,  the function  $1-\varpi_0 x/\tan x$ in the integrand is  approximated by the  Maclaurin series expansion terminated at the $x^{12}$-order term.  For small values of $x$, this should  help reduce any  ill-effect  that would arise  from the inaccuracy involved in returned values 
from  a library function  $\tan(x)$ supplied by  a  compiler  in use.
In addition, especially for $\varpi_0> 0.999$, it is crucial to first specify  the residue    $\delta=1-\varpi_0$, and subsequently calculate 
 the   value of $\varpi_0(=1-\delta)$ in order to avoid  loss of significant figures which inevitably occurs  in calculating  $1-\varpi_0$ whenever 
  $\varpi_0$ is close to unity.  \par
     In contrast, for $\varpi_0\le 10^{-3}$, the exact value of $\alpha_0(\varpi_0)$ can better be evaluated with an approximate representation like the second one of Eq.\eqref{alpha-0}, although our numerical alculations of the moments using   Eq.\eqref{mth-moment} are hardly affected under such circumstances.   
\item[5.] Accuracy of the calculated values for the $H$-function was found to  depend on the chosen values for the parameters $(\varepsilon, x_\text{t})$. A set of  optimum values for them  was  therefore searched  using  a modified Powell method for function minimization:  the quantity $\Delta^\text{Tot}(\varepsilon, x_\text{t})$ (see Eq.\eqref{deltatot}) was thereby employed as the target function. 
\end{enumerate}\par
With the optimum values obtained for $(\varepsilon, x_\text{t})$, we carried out calculations of the numerical values of  the $H$-function  for combinations of 56 values of $\varpi_0$ and 22 values of $\mu$. The results for 18 selected values of $\varpi_0$ are presented in Tables 1 through 6.
The magnitudes of  errors involved in our results  are estimated  to be at most 2  in the 15-th decimal places. In other words, our 
values of  the $H$-function for isotropic scattering $H^\text{DE}(\varpi_0, \mu)$ are supposed to be correct at least to the 14-th decimal place or   15 digits. \par 
The DE-formula is also capable of  yielding  numerical values for  the zeroth moment $\alpha_0(\varpi_0)$  to the 15-th decimal place in comparison with the exact values.
As for the first through fourth order moments $\alpha_m(\varpi_0)\quad (m=1, \cdots, 4)$, our values are obviously correct at least to the 10-th decimal place as the comparison with those of \cite{bos1983} indicates. 
It is highly likely  that they are  correct  even to the 15-th decimal  in view of the fact that ours  for $\alpha_1(1)$, $\alpha_2(1)$, and $\alpha_3(1)$ agree perfectly with the exact values and that our values of $\alpha_m(0.5)\quad(m=0, 1, 2)$ coincide with those of \cite{viik1986} to the 13-th decimal.
\vspace{1cm}\\
{\bf Acknowledgements}\quad The author is grateful to the anonymous reviewer 
for his or her valuable suggestions.


\appendix
\begin{multicols}{2}
\section{Expansion coefficients for the Hopf constant}
The Hopf constant $q_\infty$  can be calculated using the formula by Placzek and Seidel(1947):
\begin{equation}
q_\infty=\frac{6}{\pi^2}+
\frac{1}{\pi}\int_0^{\pi/2}\left(\frac{3}{x^2}
-\frac{1}{1-x\cot x}\right)dx.    \label{eq-1}
\end{equation}
Following Viik(1986), we expand the integrand of Eq.\eqref{eq-1} in a power series of $x$ up to the term of $x^n$, and carry out the integration analytically, to get the following result:
\begin{equation}
q_\infty\simeq\frac{6}{\pi^2}+\frac{1}{\pi}\sum_{n=2}^\infty\frac{(-b_n)}{2n-3}\left(\frac{\pi}{2}\right)^{2n-3},
\end{equation}
 It is found that terminating the series at $n=250$, the numerical value given by \cite{viik1986}  can be reproduced to the 59-th decimal place by rounding  off the 60-th decimal figure $8$ of our result:
 \begin{align}
   0.&7104460895987630727325241416991536719932 \notag\\
   &01333958785239092798
\end{align}   
\citep[see also][which gives  this value  to the 20-th decimal place]{lonaz2006}.\par
    We present in Table~\ref{tab-7}  the coefficients $(-b_n)$ up to $n=17$ in fraction form, ignoring all the terms beyond $(\pi/2)^{33}$, and in Table~\ref{tab-8} in decimal form. We have confirmed that the  value of $q_\infty$ can be reproduced  to the 15-th decimal figure or $0.7104460895987631$  using  a FORTRAN code in  double-precision arithmetic by incorporating  the coefficients given in  Table~\ref{tab-8}.
It must also be noted that the values for $b_n\quad (n=2, 3, 4, 5)$ shown  here are in full agreement with those cited in Viik(1986).
\end{multicols}
 \begin{table*}
 \small
 \caption{$H(\varpi_0, \mu)$ for isotropic scattering obtained by DE-formula\label{tab-1}}
 \begin{tabular}{@{}llrlrlr@{}}
 \tableline\tableline
 $\mu$ & $\varpi_0=0.001$ & $\Delta$  & $\varpi_0=0.1$ & $\Delta$ & $\varpi_0=0.3$ & $\Delta$ \\
 \hline
 $0.00$ & $1.000000000000000$ & $ 0$ & $1.000000000000000$ & $ 0$ & $ 1.000000000000000$ & $ 0$ \\
 $0.01$ & $1.000023079276825$ & $ 0$ & $1.002345867956824$ & $ 0$ & $ 1.007302106320789$ & $ 0$ \\
 $0.05$ & $1.000076131960432$ & $ 0$ & $1.007808971225585$ & $ 0$ & $ 1.024805080799005$ & $ 0$ \\
 $0.10$ & $1.000119931789617$ & $ 0$ & $1.012378097521557$ & $ 0$ & $ 1.039874888488393$ & $ 0$ \\
 $0.15$ & $1.000152819807939$ & $ 0$ & $1.015841506268183$ & $ 0$ & $ 1.051546361588496$ & $ 0$ \\
 $0.20$ & $1.000179244792271$ & $ 0$ & $1.018644075804065$ & $-1$ & $ 1.061146531525849$ & $ 0$ \\
 $0.25$ & $1.000201262444438$ & $ 0$ & $1.020992537960899$ & $ 0$ & $ 1.069298694975040$ & $ 0$ \\
 $0.30$ & $1.000220045948286$ & $ 0$ & $1.023005563463196$ & $ 0$ & $ 1.076364959992271$ & $ 0$ \\
 $0.35$ & $1.000236344202557$ & $ 0$ & $1.024759334758916$ & $ 0$ & $ 1.082580630562847$ & $ 0$ \\
 $0.40$ & $1.000250670335105$ & $ 0$ & $1.026306329929175$ & $ 0$ & $ 1.088109725497495$ & $ 1$ \\
 $0.45$ & $1.000263393666112$ & $ 0$ & $1.027684508780403$ & $ 0$ & $ 1.093072235673804$ & $ 1$ \\
 $0.50$ & $1.000274789867379$ & $ 0$ & $1.028922337487161$ & $ 0$ & $ 1.097559108679834$ & $ 0$ \\
 $0.55$ & $1.000285070583005$ & $ 0$ & $1.030041764949414$ & $ 0$ & $ 1.101641177067604$ & $ 0$ \\
 $0.60$ & $1.000294401997181$ & $ 0$ & $1.031060094515257$ & $ 0$ & $ 1.105374809040979$ & $ 0$ \\
 $0.65$ & $1.000302917028510$ & $ 0$ & $1.031991217231523$ & $ 0$ & $ 1.108805656809118$ & $ 0$ \\
 $0.70$ & $1.000310723643944$ & $ 0$ & $1.032846455396843$ & $ 0$ & $ 1.111971238341045$ & $ 0$ \\
 $0.75$ & $1.000317910704164$ & $ 0$ & $1.033635157555853$ & $ 0$ & $ 1.114902771206713$ & $ 0$ \\
 $0.80$ & $1.000324552180561$ & $ 0$ & $1.034365129091853$ & $ 0$ & $ 1.117626509039725$ & $ 0$ \\
 $0.85$ & $1.000330710264780$ & $ 0$ & $1.035042950713589$ & $-1$ & $ 1.120164736919162$ & $ 0$ \\
 $0.90$ & $1.000336437705311$ & $ 0$ & $1.035674218484590$ & $ 0$ & $ 1.122536526665526$ & $ 0$ \\
 $0.95$ & $1.000341779592330$ & $ 0$ & $1.036263727699855$ & $ 0$ & $ 1.124758319300910$ & $ 0$ \\
 $1.00$ & $1.000346774740928$ & $ 0$ & $1.036815615781196$ & $ 0$ & $ 1.126844380631286$ & $ 0$ \\
  &  & & & & & \\
 $\alpha_0$ & $1.000250125078180$ &  & $1.026334038989724$ &  & $ 1.088933156439496$ &  \\
 $\alpha_1$ & $0.500147794234295$ &  & $0.515610619015497$ &  & $ 0.553121064485779$ &  \\
 $\alpha_2$ & $0.333437556858291$ &  & $0.344358308240080$ &  & $ 0.370984157353370$ &  \\
 $\alpha_3$ & $0.250080340576093$ &  & $0.258505711234432$ &  & $ 0.279106127488061$ &  \\
 $\alpha_4$ & $0.200065314263529$ &  & $0.206918528053885$ &  & $ 0.223705287319367$ &  \\
 \tableline
 \end{tabular}
 \end{table*}

 \begin{table*}
 \small
 \caption{$H(\varpi_0, \mu)$ for isotropic scattering obtained by DE-formula\label{tab-2}}
 \begin{tabular}{@{}llrlrlr@{}}
 \tableline\tableline
 $\mu$ & $\varpi_0=0.5$ & $\Delta$  & $\varpi_0=0.7$ & $\Delta$ & $\varpi_0=0.8$ & $\Delta$ \\
 \tableline
 $0.00$ & $1.000000000000000$ & $ 0$ & $1.000000000000000$ & $ 0$ & $ 1.000000000000000$ & $ 0$ \\
 $0.01$ & $1.012723830480086$ & $ 0$ & $1.018874827015222$ & $ 0$ & $ 1.022420537254950$ & $ 0$ \\
 $0.05$ & $1.044265160581558$ & $ 0$ & $1.067654600041384$ & $ 0$ & $ 1.081914516266725$ & $ 1$ \\
 $0.10$ & $1.072368762029909$ & $ 0$ & $1.113031838677712$ & $ -1$ & $ 1.138807666285126$ & $ 0$ \\
 $0.15$ & $1.094709732081995$ & $ 0$ & $1.150343829254924$ & $ 0$ & $ 1.186640082601294$ & $ 0$ \\
 $0.20$ & $1.113461428850377$ & $ 0$ & $1.182515785241134$ & $ 0$ & $ 1.228638765535220$ & $ 0$ \\
 $0.25$ & $1.129653093499740$ & $ 0$ & $1.210934115182344$ & $ 0$ & $ 1.266322487667053$ & $ 0$ \\
 $0.30$ & $1.143889508574922$ & $ 0$ & $1.236419324639275$ & $ 0$ & $ 1.300588278431887$ & $ 0$ \\
 $0.35$ & $1.156568713007439$ & $ 0$ & $1.259517410773021$ & $ 0$ & $ 1.332034056023601$ & $ 0$ \\
 $0.40$ & $1.167971849776487$ & $ 0$ & $1.280619201798580$ & $ 0$ & $ 1.361089701705004$ & $ 0$ \\
 $0.45$ & $1.178307321633036$ & $ 0$ & $1.300018686516232$ & $ 0$ & $ 1.388080645843600$ & $ 0$ \\
 $0.50$ & $1.187735132670431$ & $ 1$ & $1.317945063118267$ & $ 0$ & $ 1.413262569404318$ & $ 0$ \\
 $0.55$ & $1.196381452566946$ & $ 0$ & $1.334581891862322$ & $ 0$ & $ 1.436842021994347$ & $ 0$ \\
 $0.60$ & $1.204347883597196$ & $ 1$ & $1.350079291216666$ & $ 0$ & $ 1.458989490017426$ & $ 0$ \\
 $0.65$ & $1.211717646238803$ & $ 0$ & $1.364562099754409$ & $ 0$ & $ 1.479848112148857$ & $ 1$ \\
 $0.70$ & $1.218559869097970$ & $ 0$ & $1.378135560424163$ & $ 0$ & $ 1.499539735979634$ & $ 0$ \\
 $0.75$ & $1.224932658619432$ & $ 0$ & $1.390889410611728$ & $ 0$ & $ 1.518169271958706$ & $ 0$ \\
 $0.80$ & $1.230885353499438$ & $ 0$ & $1.402900906035831$ & $ 0$ & $ 1.535827913161673$ & $ 0$ \\
 $0.85$ & $1.236460217044641$ & $ 0$ & $1.414237108074479$ & $ 0$ & $ 1.552595573997779$ & $ 0$ \\
 $0.90$ & $1.241693731628014$ & $ 0$ & $1.424956647948121$ & $ 0$ & $ 1.568542775461757$ & $ 0$ \\
 $0.95$ & $1.246617604949040$ & $ 0$ & $1.435111110389834$ & $ 0$ & $ 1.583732128410658$ & $ 0$ \\
 $1.00$ & $1.251259563383223$ & $ 1$ & $1.444746134765130$ & $ 0$ & $ 1.598219518533160$ & $ 1$ \\
  &  & & & & & \\
 $\alpha_0$ & $1.171572875253810$ &  & $1.292221264270954$ &  & $ 1.381966011250105$ &  \\
 $\alpha_1$ & $0.603484255848994$ &  & $0.678667819110035$ &  & $ 0.735815233031298$ &  \\
 $\alpha_2$ & $0.407023642126835$ &  & $0.461419889227813$ &  & $ 0.503223793206828$ &  \\
 $\alpha_3$ & $0.307119523431517$ &  & $0.349675124789374$ &  & $ 0.382595118839524$ &  \\
 $\alpha_4$ & $0.246600776618178$ &  & $0.281528145681548$ &  & $ 0.308664834854489$ &  \\
 \tableline
 \end{tabular}
 \end{table*}
 \begin{table*}
 \small
 \caption{$H(\varpi_0, \mu)$ for isotropic scattering obtained by DE-formula\label{tab-3}}
 \begin{tabular}{@{}llrlrlr@{}}
 \tableline\tableline
 $\mu$ & $\varpi_0=0.9$ & $\Delta$  & $\varpi_0=0.99$ & $\Delta$ & $\varpi_0=0.999$ & $\Delta$ \\
 \tableline
 $0.00$ & $1.000000000000000$ & $ 0$ & $1.000000000000000$ & $ 0$ & $ 1.000000000000000$ & $ 0$ \\
 $0.01$ & $1.026603764674318$ & $ 0$ & $1.032257470361386$ & $ 0$ & $ 1.033674233753012$ & $ 0$ \\
 $0.05$ & $1.099678288027236$ & $ 1$ & $1.126070564855184$ & $ 0$ & $ 1.133393124789600$ & $ 0$ \\
 $0.10$ & $1.172143053516453$ & $ 0$ & $1.224875290417156$ & $ 0$ & $ 1.240424772856097$ & $ -1$ \\
 $0.15$ & $1.234918332479768$ & $ 0$ & $1.314972472230572$ & $ 0$ & $ 1.339648497723789$ & $ 0$ \\
 $0.20$ & $1.291433728065182$ & $ 0$ & $1.399767224074789$ & $ 0$ & $ 1.434415855969377$ & $ 0$ \\
 $0.25$ & $1.343270825469927$ & $ 0$ & $1.480742606314496$ & $ 1$ & $ 1.526164843765312$ & $ 0$ \\
 $0.30$ & $1.391350320632572$ & $ 0$ & $1.558712877184762$ & $ 0$ & $ 1.615671055040806$ & $ 1$ \\
 $0.35$ & $1.436280343349813$ & $ 0$ & $1.634185389649200$ & $ 1$ & $ 1.703407663333851$ & $ 0$ \\
 $0.40$ & $1.478496236155405$ & $ 0$ & $1.707503465981193$ & $ 0$ & $ 1.789687085591210$ & $ 0$ \\
 $0.45$ & $1.518327492340675$ & $ 1$ & $1.778913614217836$ & $ 1$ & $ 1.874727429567537$ & $1$ \\
 $0.50$ & $1.556033802021363$ & $ 0$ & $1.848601016447846$ & $ 1$ & $ 1.958687474359325$ & $ 0$ \\
 $0.55$ & $1.591826182604205$ & $ 0$ & $1.916709878377813$ & $ 0$ & $ 2.041686670750376$ & $ 0$ \\
 $0.60$ & $1.625880196208851$ & $ 0$ & $1.983355849810225$ & $ 0$ & $ 2.123817312270080$ & $ 0$ \\
 $0.65$ & $1.658344654834039$ & $ 0$ & $2.048633986828251$ & $ 0$ & $ 2.205152313093988$ & $ 0$ \\
 $0.70$ & $1.689347600330646$ & $ 1$ & $2.112624062808928$ & $ 1$ & $ 2.285750378701103$ & $ 0$ \\
 $0.75$ & $1.719000559169012$ & $ 0$ & $2.175394229576233$ & $ 1$ & $ 2.365659557180583$ & $ -1$ \\
 $0.80$ & $1.747401661141401$ & $ 0$ & $2.237003612379969$ & $ 0$ & $ 2.444919746120597$ & $ 0$ \\
 $0.85$ & $1.774637984393825$ & $ 1$ & $2.297504193613621$ & $ 0$ & $ 2.523564504125438$ & $ 0$ \\
 $0.90$ & $1.800787358056601$ & $ 0$ & $2.356942208926965$ & $ 0$ & $ 2.601622386587422$ & $ 0$ \\
 $0.95$ & $1.825919774834691$ & $ 0$ & $2.415359201062581$ & $ 1$ & $ 2.679117948214393$ & $ 0$ \\
 $1.00$ & $1.850098516769812$ & $ -1$ & $2.472792828397026$ & $ 0$ & $ 2.756072507268736$ & $ 0$ \\
  &  & & & & & \\
 $\alpha_0$ & $1.519493853295915$ &  & $1.818181818181818$ &  & $ 1.938693139936569$ &  \\
 $\alpha_1$ & $0.825315748434260$ &  & $1.027181735489862$ &  & $ 1.111330550198365$ &  \\
 $\alpha_2$ & $0.569448617885390$ &  & $0.721955096966521$ &  & $ 0.786717738091024$ &  \\
 $\alpha_3$ & $0.435113601428221$ &  & $0.557657562330954$ &  & $ 0.610325627802045$ &  \\
 $\alpha_4$ & $0.352161961828424$ &  & $0.454579895752381$ &  & $ 0.498970970855085$ &  \\
 \tableline
 \end{tabular}
 \end{table*}
 \begin{table*}
 \small
 \caption{$H(\varpi_0, \mu)$ for isotropic scattering obtained by DE-formula\label{tab-4}}
 \begin{tabular}{@{}llrlrlr@{}}
 \tableline\tableline
 $\mu$ & $\varpi_0=1-10^{-5}$ & $\Delta$  & $\varpi_0=1-10^{-7}$ & $\Delta$ & $\varpi_0=1-10^{-9}$ & $\Delta$ \\
 \tableline
 $0.00$ & $1.000000000000000$ & $ 0$ & $1.000000000000000$ & $ 0$ & $ 1.000000000000000$ & $ 0$ \\
 $0.01$ & $1.034205719689598$ & $ 0$ & $1.034256922275328$ & $ 0$ & $ 1.034262022863829$ & $ 1$ \\
 $0.05$ & $1.136262879388996$ & $ 0$ & $1.136543713405656$ & $ 0$ & $ 1.136571734129853$ & $ 0$ \\
 $0.10$ & $1.246666275155322$ & $ 0$ & $1.247282112614023$ & $ 0$ & $ 1.247343610377832$ & $ 0$ \\
 $0.15$ & $1.349722855408363$ & $ 0$ & $1.350722601325096$ & $ 0$ & $ 1.350822494497537$ & $ 0$ \\
 $0.20$ & $1.448762094088692$ & $ 0$ & $1.450192529327167$ & $ 0$ & $ 1.450335524951871$ & $ 0$ \\
 $0.25$ & $1.545207661341358$ & $ 0$ & $1.547114359446071$ & $ 0$ & $ 1.547305046360939$ & $ 0$ \\
 $0.30$ & $1.639824632440255$ & $ 0$ & $1.642252383517436$ & $ 0$ & $ 1.642495275204116$ & $ 0$ \\
 $0.35$ & $1.733077835787557$ & $ 0$ & $1.736070880193769$ & $ 0$ & $ 1.736370438341392$ & $ 1$ \\
 $0.40$ & $1.825272706424626$ & $ 0$ & $1.828874877403022$ & $ -1$ & $ 1.829235526266891$ & $ 0$ \\
 $0.45$ & $1.916621283338394$ & $ 0$ & $1.920876100191321$ & $ 0$ & $ 1.921302235960664$ & $ 1$ \\
 $0.50$ & $2.007276919589851$ & $ 0$ & $2.012227651599931$ & $ 1$ & $ 2.012723648818164$ & $ 1$ \\
 $0.55$ & $2.097354110520160$ & $ 0$ & $2.103043822474229$ & $ 0$ & $ 2.103614038607345$ & $ 1$ \\
 $0.60$ & $2.186940550203905$ & $ 0$ & $2.193412135789794$ & $ 0$ & $ 2.194060914574986$ & $ 0$ \\
 $0.65$ & $2.276104830260681$ & $ 0$ & $2.283401037166935$ & $ 1$ & $ 2.284132711127234$ & $ 0$ \\
 $0.70$ & $2.364901554295879$ & $ 1$ & $2.373065003460309$ & $ 1$ & $ 2.373883895831389$ & $ 1$ \\
 $0.75$ & $2.453374848241580$ & $-1$ & $2.462448048868194$ & $ 1$ & $ 2.463358475100236$ & $ 0$ \\
 $0.80$ & $2.541560836748558$ & $ 0$ & $2.551586198198066$ & $ 0$ & $ 2.552592467143796$ & $ 1$ \\
 $0.85$ & $2.629489431573132$ & $-1$ & $2.640509272900804$ & $ 0$ & $ 2.641615687767435$ & $ 0$ \\
 $0.90$ & $2.717185649514410$ & $ 0$ & $2.729242207208971$ & $ 1$ & $ 2.730453066329686$ & $ 0$ \\
 $0.95$ & $2.804670600985655$ & $ 1$ & $2.817806035313179$ & $ 2$ & $ 2.819125632778694$ & $ 0$ \\
 $1.00$ & $2.891962243185030$ & $-2$ & $2.906218643437432$ & $ 0$ & $ 2.907651269617728$ & $ -1$ \\
  &  & & & & & \\
 $\alpha_0$ & $1.993695381633480$ &  & $1.999367744404741$ &  & $ 1.999936756446733$ &  \\
 $\alpha_1$ & $1.150223392945239$ &  & $1.154251374421934$ &  & $ 1.154655607439818$ &  \\
 $\alpha_2$ & $0.816872433717816$ &  & $0.820003269398072$ &  & $ 0.820317548751288$ &  \\
 $\alpha_3$ & $0.634969398790733$ &  & $0.637532347623988$ &  & $ 0.637789665617137$ &  \\
 $\alpha_4$ & $0.519814783449839$ &  & $0.521985149313924$ &  & $ 0.522203079284880$ &  \\
 \tableline
 \end{tabular}
 \end{table*}
 \begin{table*}
 \small
 \caption{$H(\varpi_0, \mu)$ for isotropic scattering obtained by DE-formula\label{tab-5}}
 \begin{tabular}{@{}llrlrlr@{}}
\tableline\tableline
  $\mu$ & $\varpi_0=1-10^{-10}$ & $\Delta$  & $\varpi_0=1-10^{-11}$ & $\Delta$ & $\varpi_0=1-10^{-12}$ & $\Delta$ \\
 \tableline
 $0.00$ & $1.000000000000000$ & $ 0$ & $1.000000000000000$ & $ 0$ & $ 1.000000000000000$ & $ 0$ \\
 $0.01$ & $1.034262410233136$ & $ 0$ & $1.034262532725766$ & $ 0$ & $ 1.034262571460906$ & $ 0$ \\
 $0.05$ & $1.136573862529025$ & $ 1$ & $1.136574535574220$ & $ 0$ & $ 1.136574748408426$ & $ 0$ \\
 $0.10$ & $1.247348282010408$ & $ -1$ & $1.247349759291493$ & $ 0$ & $ 1.247350226446905$ & $ 0$ \\
 $0.15$ & $1.350830083242122$ & $ 0$ & $1.350832482995976$ & $ 0$ & $ 1.350833241862986$ & $ 0$ \\
 $0.20$ & $1.450346388639971$ & $ 1$ & $1.450349824029179$ & $ 0$ & $ 1.450350910393573$ & $ 0$ \\
 $0.25$ & $1.547319533862394$ & $ 0$ & $1.547324115216175$ & $ 0$ & $ 1.547325563967803$ & $ 0$ \\
 $0.30$ & $1.642513729686032$ & $ 0$ & $1.642519565530900$ & $ 0$ & $ 1.642521410989613$ & $ 0$ \\
 $0.35$ & $1.736393199059862$ & $ 1$ & $1.736400396686262$ & $ 0$ & $ 1.736402672781104$ & $ 1$ \\
 $0.40$ & $1.829262929658623$ & $ -1$ & $1.829271595466160$ & $ 0$ & $ 1.829274335844537$ & $ 0$ \\
 $0.45$ & $1.921334616355098$ & $ 0$ & $1.921344856077736$ & $ 0$ & $ 1.921348094176634$ & $ 0$ \\
 $0.50$ & $2.012761338925487$ & $ 0$ & $2.012773257785881$ & $ 1$ & $ 2.012777026880672$ & $ 1$ \\
 $0.55$ & $2.103657369865718$ & $ 1$ & $2.103671072685019$ & $ 0$ & $ 2.103675405924182$ & $ 0$ \\
 $0.60$ & $2.194110217405046$ & $ 0$ & $2.194125808683393$ & $ 0$ & $ 2.194130739113968$ & $ -1$ \\
 $0.65$ & $2.284188315123211$ & $ 0$ & $2.284205899100158$ & $ 1$ & $ 2.284211459686857$ & $ 0$ \\
 $0.70$ & $2.373946129907529$ & $ 0$ & $2.373965810608357$ & $ 0$ & $ 2.373972034248210$ & $ 0$ \\
 $0.75$ & $2.463427667604941$ & $ 0$ & $2.463449548876749$ & $ 1$ & $ 2.463456468410514$ & $ 0$ \\
 $0.80$ & $2.552668945949899$ & $ 0$ & $2.552693131490155$ & $ -1$ & $ 2.552700779711349$ & $ 0$ \\
 $0.85$ & $2.641699780344541$ & $ 0$ & $2.641726373723957$ & $ 1$ & $ 2.641734783386086$ & $ 0$ \\
 $0.90$ & $2.730545099802790$ & $ 0$ & $2.730574204483862$ & $ 1$ & $ 2.730583408306289$ & $ 0$ \\
 $0.95$ & $2.819225933976175$ & $ 0$ & $2.819257653328354$ & $ 0$ & $ 2.819267684001096$ & $ 1$ \\
 $1.00$ & $2.907760165110989$ & $ 0$ & $2.907794602423222$ & $ -1$ & $ 2.907805492610912$ & $ 0$ \\
  &  & & & & & \\
 $\alpha_0$ & $1.999980000199998$ &  & $1.999993675464680$ &  & $ 1.999998000002000$ &  \\
 $\alpha_1$ & $1.154686329619116$ &  & $1.154696045139821$ &  & $ 1.154699117488689$ &  \\
 $\alpha_2$ & $0.820341434947413$ &  & $0.820348988688071$ &  & $ 0.820351377416827$ &  \\
 $\alpha_3$ & $0.637809222904595$ &  & $0.637815407686305$ &  & $ 0.637817363508448$ &  \\
 $\alpha_4$ & $0.522219643112460$ &  & $0.522224881250612$ &  & $ 0.522226537714934$ &  \\
 \tableline
 \end{tabular}
 \end{table*}
 \begin{table*}
 \small
 \caption{$H(\varpi_0, \mu)$ for isotropic scattering obtained by DE-formula\label{tab-6}}
 \begin{tabular}{@{}llrlrlr@{}}
 \tableline\tableline
 $\mu$ & $\varpi_0=1-10^{-13}$ & $\Delta$  & $\varpi_0=1-10^{-14}$ & $\Delta$ & $\varpi_0=1$ & $\Delta$ \\
 \tableline
 $0.00$ & $1.000000000000000$ & $ 0$ & $1.000000000000000$ & $ 0$ & $ 1.000000000000000$ & $ 0$ \\
 $0.01$ & $1.034262583709990$ & $ 0$ & $1.034262587583486$ & $ 0$ & $ 1.034262589374882$ & $ 0$ \\
 $0.05$ & $1.136574815712375$ & $ 0$ & $1.136574836995738$ & $ 0$ & $ 1.136574846838766$ & $ 0$ \\
 $0.10$ & $1.247350374174229$ & $ 0$ & $1.247350420889692$ & $ 0$ & $ 1.247350442494436$ & $ 0$ \\
 $0.15$ & $1.350833481837627$ & $ 0$ & $1.350833557724253$ & $ 0$ & $ 1.350833592819941$ & $ 1$ \\
 $0.20$ & $1.450351253932052$ & $ 0$ & $1.450351362568447$ & $ -1$ & $ 1.450351412810095$ & $ 0$ \\
 $0.25$ & $1.547326022103329$ & $ 0$ & $1.547326166978507$ & $ 0$ & $ 1.547326233979698$ & $ 0$ \\
 $0.30$ & $1.642521994575152$ & $ 0$ & $1.642522179121128$ & $ 0$ & $ 1.642522264469087$ & $ -1$ \\
 $0.35$ & $1.736403392546043$ & $ 0$ & $1.736403620155757$ & $ 0$ & $ 1.736403725419636$ & $ 0$ \\
 $0.40$ & $1.829275202429211$ & $-1$ & $1.829275476467441$ & $ 0$ & $ 1.829275603203367$ & $-1$ \\
 $0.45$ & $1.921349118154844$ & $ 0$ & $1.921349441965328$ & $ 0$ & $ 1.921349591719701$ & $ 0$ \\
 $0.50$ & $2.012778218775117$ & $ 0$ & $2.012778595685436$ & $ 0$ & $ 2.012778769997181$ & $ 0$ \\
 $0.55$ & $2.103676776217445$ & $ -1$ & $2.103677209542494$ & $-1$ & $ 2.103677409944670$ & $ 0$ \\
 $0.60$ & $2.194132298256560$ & $ 0$ & $2.194132791301094$ & $ 1$ & $ 2.194133019322067$ & $ 0$ \\
 $0.65$ & $2.284213218103263$ & $ 0$ & $2.284213774163804$ & $ 0$ & $ 2.284214031328140$ & $ 0$ \\
 $0.70$ & $2.373974002341517$ & $ -1$ & $2.373974624707825$ & $ 0$ & $ 2.373974912536958$ & $ 0$ \\
 $0.75$ & $2.463458656566025$ & $ 0$ & $2.463459348522235$ & $ 1$ & $ 2.463459668534998$ & $ 0$ \\
 $0.80$ & $2.552703198299433$ & $ 0$ & $2.552703963124959$ & $ 0$ & $ 2.552704316838003$ & $ 1$ \\
 $0.85$ & $2.641737442764470$ & $ 0$ & $2.641738283734728$ & $ 1$ & $ 2.641738672662854$ & $ 0$ \\
 $0.90$ & $2.730586318821909$ & $ 0$ & $2.730587239208903$ & $-1$ & $ 2.730587664865336$ & $ 0$ \\
 $0.95$ & $2.819270855991615$ & $ 1$ & $2.819271859064420$ & $ 1$ & $ 2.819272322961027$ & $ 0$ \\
 $1.00$ & $2.907808936405973$ & $-1$ & $2.907810025431127$ & $ 0$ & $ 2.907810529078606$ & $ 0$ \\
  &  & & & & & \\
 $\alpha_0$ & $1.999999367544668$ &  & $1.999999800000020$ &  & $ 2.000000000000000$ &  \\
 $\alpha_1$ & $1.154700089053853$ &  & $1.154700396290050$ &  & $ 1.154700538379251$ &  \\
 $\alpha_2$ & $0.820352132801808$ &  & $0.820352371675775$ &  & $ 0.820352482149125$ &  \\
 $\alpha_3$ & $0.637817981995959$ &  & $0.637818177579107$ &  & $ 0.637818268031518$ &  \\
 $\alpha_4$ & $0.522227061536906$ &  & $0.522227227184154$ &  & $ 0.522227303791946$ &  \\
 \tableline
 \end{tabular}
 \end{table*}
\begin{table*}
\small
\caption{Expansion coefficients for the Hopf constant $q_\infty$ in fraction form\label{tab-7}}
\begin{tabular}{ccl}\\ \tableline\tableline
$n$ &  $2n-3$ &  $-b_n$  \\ \tableline
2    & 1 & $1/5$  \\
3    &  3 & $1/175$ \\
4    &  5 & $2/7875$ \\
5    &  7 & $37/3031875$\\   
6    &  9  & $118/197071875$ \\
7    &  11 & $5506/186232921875$ \\
8    &  13  & $3308/2261399765625$\\
9    &  15  & $8386459 /115794974998828125$\\
10   & 17  & $56067454/15632321624841796875$\\
11   & 19 & $29059045766/163592245803969404296875$\\
12   &  21 & $654185668/74360111729077001953125$\\
13   &  23  & $94994770034/218018127581059225341796875$ \\
14   &  25  & $682205256508/31612628499253587674560546875$\\
15   &  27  & $8101862427463268/7580234124693521520544500732421875$\\
16   &  29  & $2112667618521098312/39909932666511390805666796356201171875$\\
17   &  31  & $765408870503963716543/291941157455530823743452615345611572265625$\\ 
\tableline
\end{tabular}
\end{table*}
\begin{table*}
\small
\caption{Expansion coefficients for the Hopf constant $q_\infty$ in decimal form\label{tab-8} } 
\begin{tabular}{ll|ll} \\ \hline\hline
  $n$   & $-b_n$      &  $n$ &  $-b_n$   \\  \hline         
   1 & $3$                               &  11 &  $1.776309483569356(-13)$ \\ 
   2 & $2.000000000000000(-1)^\text{a}$ &  12 &  $8.797534764114590(-15)$  \\
   3 & $5.714285714285714(-3)$ &  13 &  $4.357195939988105(-16)$\\
   4 & $2.539682539682540(-4)$ &  14 &  $2.158014973427811(-17)$ \\
   5 & $1.220366934652649(-5)$ &  15 &  $1.068814273304630(-18)$ \\
   6 & $5.987663130520273(-7)$ &  16 &  $5.293588531392957(-20)$ \\
   7&  $2.956512707079601(-8)$  &  17 &  $2.621791586958932(-21)$ \\  
   8 & $1.462810799878960(-9)$ &      &                        \\
   9 & $7.242506853242011(-11)$  &    &                        \\
   10 & $3.586636415598148(-12)$ &   &                        \\      \hline
\end{tabular}
\tablenotetext{\text{a}}{This is to read as $2.000000000000000\times 10^{-1}$.}
\end{table*}
\end{document}